\documentclass[
 aip,
 amsmath,amssymb,
 reprint,
]{revtex4-1}

\usepackage{graphicx}
\usepackage{dcolumn}
\usepackage{bm}
\usepackage{color}

\usepackage[utf8]{inputenc}
\usepackage[T1]{fontenc}
\usepackage{mathptmx}
\usepackage{natbib}

\begin{document}

\preprint{AIP/123-QED}

\title{Three-Omega Thermal-Conductivity Measurements with Curved Heater Geometries}
 
\author{Gabriel R. Jaffe}
\author{Keenan J. Smith}
\author{Victor W. Brar}
\affiliation{Department of Physics, University of Wisconsin-Madison, Madison, Wisconsin 53706, USA}

\author{Max G. Lagally}
\affiliation{Department of Materials Science and Engineering, University of Wisconsin-Madison, Madison, Wisconsin 53706, USA}

\author{Mark A. Eriksson}
\affiliation{Department of Physics, University of Wisconsin-Madison, Madison, Wisconsin 53706, USA}

\date{\today}

\begin{abstract}
The three-omega method, a powerful technique to measure the thermal conductivity of nanometer-thick films and the interfaces between them, has historically employed straight conductive wires to act as both heaters and thermometers.  When investigating stochastically prepared samples such as two-dimensional materials and nanomembranes, residue and excess material can make it difficult to fit the required millimeter-long straight wire on the sample surface.  There are currently no available criteria for how diverting three-omega heater wires around obstacles affects the validity of the thermal measurement.  In this Letter, we quantify the effect of wire curvature by performing three-omega experiments with a wide range of frequencies using both curved and straight heater geometries on SiO$_2$/Si samples. When the heating wire is curved, we find that the measured Si substrate thermal conductivity changes by only 0.2\%. Similarly, we find that wire curvature has no significant effect on the determination of the thermal resistance of a $\sim$65\,nm SiO$_2$ layer, even for the sharpest corners considered here, for which the largest measured ratio of the thermal penetration depth of the applied thermal wave to radius of curvature of the heating wire is 4.3. This result provides useful design criteria for three-omega experiments by setting a lower bound for the maximum ratio of thermal penetration depth to wire radius of curvature.\end{abstract}

\maketitle

In recent years two-dimensional (2D) materials and nanomembranes (NMs) have garnered significant interest for their novel thermal\cite{Balandin_NATMAT_2011,Luo_NATCOM_2015,Kang_SCI_2018,Guo_MatSciEngRep_2018}, electronic\cite{Bonaccorso_SCI_2015,Novoselovaac_SCI_2016,Tan_ChemRev_2017,Yuan_NAT_2018,Guo_MatSciEngRep_2018}, and optical properties\cite{Novoselovaac_SCI_2016,Tan_ChemRev_2017,Guo_MatSciEngRep_2018}.  Precise thermal-conductivity measurements of both the thin layers and the interfaces present in these samples are crucial for advancing our understanding of thermal transport and informing thermal-management efforts in devices.\cite{Cahill_JAP_2003,Pop_NR_2010} The three-omega method is a well established technique for measuring the thermal conductivity of thin-films and interfaces.\cite{Cahill_RSI_1990,Lee_JAP_1997,Kim_APL_2000,Kim_PRL_2006,Chen_APL_2009,Schroeder_PRL_2015,Yang_AFM_2018,Jaffe_ACSAMI_2019,Xu_JAP_2019,Velarde_ACEAMI_2019} This technique utilizes a conductive four-probe wire as both a heater and thermometer and is able to measure simultaneously the thermal conductivity of a thin-film and the substrate beneath it.  A central assumption in the three-omega method is that the heating wire, which is typically around one millimeter in length, acts as a straight infinite-line source of heat.\cite{Cahill_RSI_1990}   Sample preparation of 2D materials and NMs is, in many cases, stochastic in nature and leaves the films of interest surrounded by undesired residue, wrinkles, and excess material.\cite{Cavallo_SoftMat_2010,Rogers_NAT_2011,Pizzocchero_NC_2016}  These obstacles present challenges for three-omega experiments, because they prevent the fabrication of straight, millimeter-long wires.  It is therefore of critical importance to understand how curved a heating wire can be and still be acceptable for thermal measurements.

In this Letter, we perform three-omega experiments with both straight and curved heater geometries on SiO$_2$ films of two different thickness supported by Si substrates.  We find that measurements using wires with radii of curvature down to 200\,$\mu$m are just as accurate as straight wires.  The curvature of the wire does not appear to affect the measurement, even when the thermal penetration depth into the substrate, which determines the sensitivity of the experiment to nonuniformity in the wire geometry, is more than four times larger than the minimum radius of curvature of the wire path.  On average, the measured cross-plane thermal-resistance difference between a 220\,nm and a 285\,nm thick SiO$_2$ film with straight and curved heater geometries differs by only 4.3\,m$^2$KGW$^{-1}$, and the measured Si substrate thermal-conductivities differ by 0.2 Wm$^{-1}$K$^{-1}$.  

Three-omega experiments are most sensitive to the thermal properties of the surrounding material in a cylindrical half volume with radius equal to the penetration depth of the thermal wave emitted by the heating wire. An ac current, passed through the heating wire at frequency $\omega$, generates a thermal wave with frequency 2$\omega$.  The thermal penetration depth $\lambda$ into the cylindrical half volume is
\begin{equation}
\lambda = \sqrt{\frac{D}{2\omega}},
\end{equation}
where $D$ is the thermal diffusivity of the supporting material\cite{Cahill_RSI_1990}.  It is important to compare the length scale of the wire curvature with the sensitivity of the measurement.  There is currently no metric for how significant an effect wire curvature has on a three-omega thin-film thermal-conductivity measurement and how any associated errors scale with $\lambda$. 

 In a three-omega measurement of a thin-film covering a substrate, the measured thermal resistance is the series sum of the thermal-resistance contributions from the substrate and the thin-film.  The substrate contribution varies linearly with the natural logarithm of the heating frequency, and the thin-film contribution is independent of heating frequency, provided that the heater is much wider than the film thickness.  In order to determine the thermal resistance of a thin-film, a differential experiment is performed where the contribution of the substrate is removed by subtracting the thermal resistance measured from a heater on the substrate from one positioned on the film of interest\cite{Cahill_APL_1994}.  If the reference and thin-film heater geometries are not identical, deviations in the measured thermal resistance from geometrical factors such as wire curvature would also affect the determination of the thin-film thermal resistance. 

\begin{figure}[t]
	\centering
	\includegraphics[width = \columnwidth]{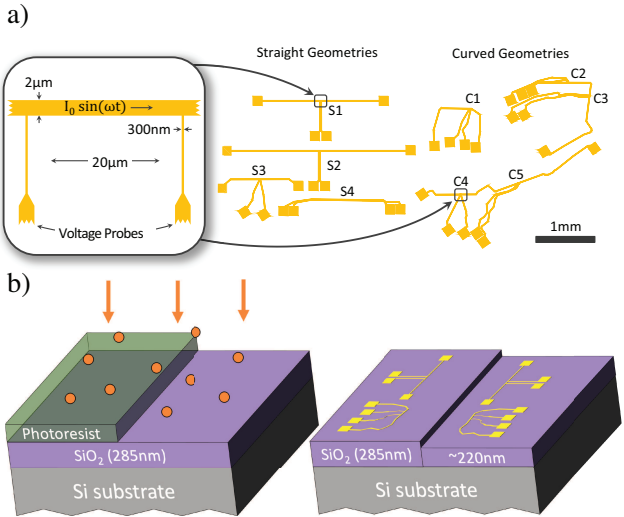}
	\caption{a) Schematic diagram of the four-probe three-omega heater geometries tested in this experiment.  All heater designs have identical widths and voltage probe spacings as seen in the popout. b) CHF$_3$ plasma etching of SiO$_2$ using a photoresist mask and patterning of three-omega heaters using e-beam lithography.}
	\label{fig:1}
\end{figure}

 The heater wire geometries tested here (Fig.\,\ref{fig:1}a), are identical to designs used by us on samples where complex routing of the heating wires is necessary.  The SiO$_2$/Si substrates in this work serve as test films with well understood thermal properties that enable us to quantify what effect the heater wire geometry has on thermal measurements.  Samples are prepared by dicing 8x8mm dies from commercially purchased Si wafers with a 285\,nm thermally grown SiO$_2$ layer.  The Si substrates are 500\,$\mu$m thick and have an electrical resistivity of $\sim$0.01\,$\Omega$\,cm.  A step is etched into the SiO$_2$ film using a CHF$_3$ plasma and a photoresist mask, as depicted in Fig.\,\ref{fig:1}b.  The height of the step is measured with an atomic force microscope to be 64.6~$\pm$~0.7\,nm.  Pairs of metal heaters of identical shapes are fabricated on each side of the step using electron beam lithography and a metal liftoff process.  All wires are 70\,nm thick (65\,nm Au with a 5\,nm Ti adhesion layer).	

Differential three-omega experiments are performed on these samples in a temperature-controlled chamber at 22\,C.  The heaters on the etched side of the SiO$_2$ films are used as the reference heaters.  For all heater geometries, current is passed between the outer bond pads and the inner bond pads are used as voltage probes to measure the temperature of the wire.  The heating wire width is maintained at 2\,$\text{$\mu$}$m for the entire length between the outer bond pads.  The section of heating wire between the voltage probes is always straight and 20\,$\mu$m long, as seen in the popout of Fig.\,\ref{fig:1}a.  Measurements are carried out on four heaters simultaneously to prevent systematic errors from drift in the ambient conditions for the measurement electronics.

\begin{figure}[b]
	\centering
	\includegraphics[width = 14 cm]{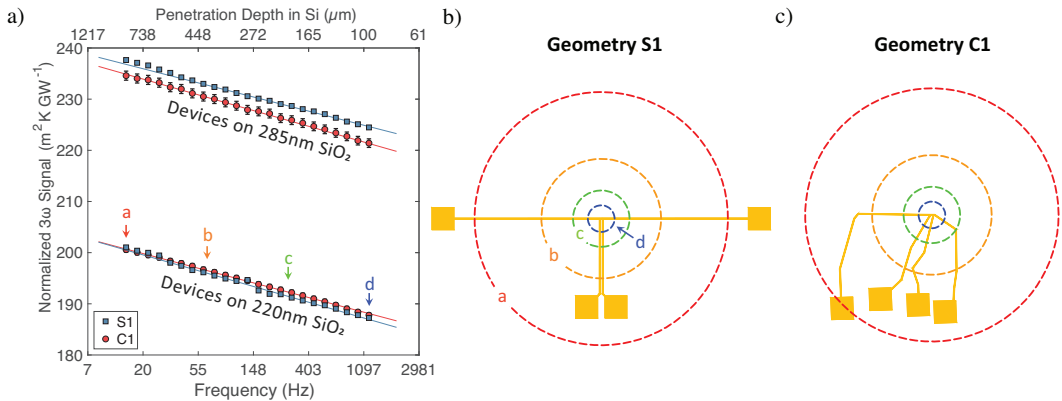}
	\caption{a) The normalized 3$\omega$ signal measured with heater geometries S1 (blue) and C1 (red) on two different thicknesses of SiO$_2$ films on Si substrates as a function of heating frequency (bottom axes) and the corresponding thermal penetration depth in Si (top axes).  The bottom x-axis is plotted on a natural logarithm scale.  b) and c) show heater schematic diagrams S1 and C1 respectively, overlaid with circles whose radii represent a range of penetration depths (indicated in the plot with colored letters).  The circles are centered between the voltage probes where the temperature rise is measured.}
	\label{fig:2}
\end{figure}

Figure\,\ref{fig:2}a shows the normalized three-omega signal as a function of heating frequency 2$\omega$ for heater geometries S1 (blue squares) and C1 (red circles) on both thicknesses of the SiO$_2$ films.  The effective thermal penetration depth in Si, using the Si thermal conductivity measured in this work, is plotted on the top x-axis.  We observe the predicted linear trend of the normalized three-omega signal with the natural logarithm of the heating frequency for the entire range measured.\cite{Cahill_RSI_1990}  To better visualize how the sensitivity of the experiment compares with the heater geometry, circles, centered between the voltage probes with radii equal to a range of penetration depths, are overlaid on the heater geometry S1 in Fig.\,\ref{fig:2}b and geometry C1 in Fig.\,\ref{fig:2}c.  There is good agreement of the measured thermal resistance between sample geometries over the entire range of applied heating frequencies.

\onecolumngrid

\newpage
\twocolumngrid


\onecolumngrid

\begin{figure}[h]
	\centering
	\includegraphics[width=14cm]{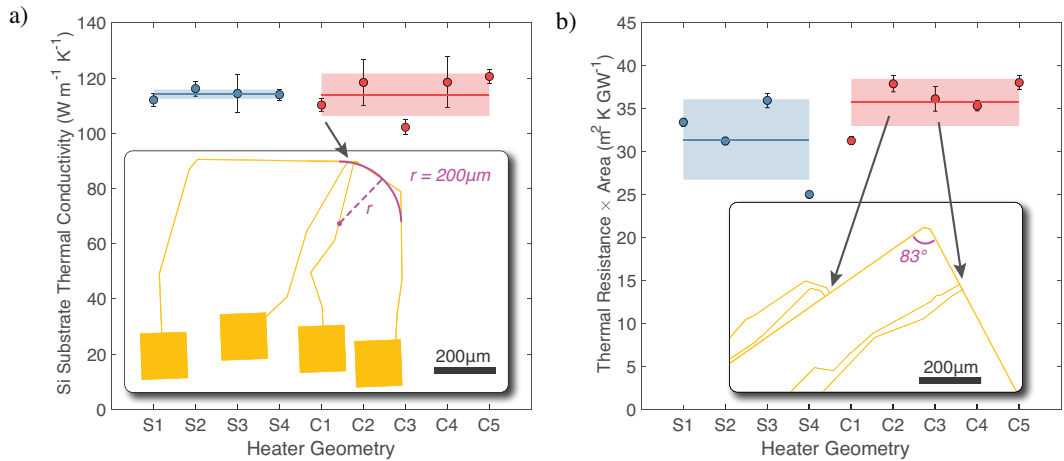}
	\caption{a) Thermal conductivity of the Si substrate as measured by straight (blue) and curved (red) heater geometries at 22\,C.  b) Cross-plane thermal-resistance difference between 285\,nm SiO$_2$ and 220\,nm SiO$_2$ films as measured by straight (blue) and curved (red) heater geometries.  The heater geometry is labeled on the x-axis.  The solid lines and shaded regions indicate the mean and one standard deviation respectively of the data with straight and curved heaters.  The insets show the radii of curvature of the heater wires near where the temperature of the wire is measured for a few select points.}
	\label{fig:3}
\end{figure}
\twocolumngrid


The three-omega method relies on an approximate solution to the heat equation that assumes heat is dissipating from a finite width and infinitely long straight-wire on the surface of an infinite half-volume\cite{Cahill_RSI_1990}.  Curved heating wires must eventually break the assumed radial symmetry of this solution once the thermal penetration depth is significantly larger than the wire radius of curvature.  The largest ratio of penetration depth to wire radius of curvature measured here is 4.3, for which we find no detectable difference in the three-omega signal when compared with measurements using straight wires.  At some larger ratio, the three-omega signal of a curved wire must begin to diverge from that of a straight wire.  In this experiment, the thermal penetration depth was limited to $\sim$1mm by the finite size of the sample die, the length of the heating wires, and lateral spacing of the heaters, which ultimately limited the maximum ratio of penetration depth to wire radius of curvature that could be tested on the given wire geometries.

The thermal conductivity of the Si substrate as measured by each unique heater geometry, is plotted in Fig.\,\ref{fig:3}a.  Measurements using straight heaters are plotted in blue, the curved are in red, and the geometry is labeled on the x-axis.  The lines and shaded squares denote the average and one standard deviation respectively for the straight and curved heaters.  We find that the Si substrate thermal conductivity measured with curved heater devices differs by only 0.2\% from the conductivity measured with straight heaters.  The average Si thermal conductivity across all geometries is 114 Wm$^{-1}$K$^{-1}$, which is in agreement with previous studies of doped Si with similar electrical conductivities.\cite{Slack_JAP_1964}

The measured thermal-resistance difference between the 220\,nm and the 285\,nm SiO$_2$ films for each heater geometry is shown in Fig.\ \ref{fig:3}b.  We find no significant difference in the resistance measured with curved or straight geometries to within the measurement uncertainty.  The average resistance measured with curved heaters is 4.3\,$\pm$\,5.4\,m$^2$K\,GW$^{-1}$ larger than the resistance measured with straight heaters.  The relative spread in the SiO$_2$ resistance data is larger than the Si substrate conductivity data because the thermal-resistance contribution of $\sim$65\,nm of SiO$_2$ constitutes only 16\% of the total thermal resistance measured by a heater.  

The thermal conductivity of the SiO$_2$ film can be determined by dividing the thickness difference between the two films by the thermal-resistance difference, assuming the trend in thermal resistance of SiO$_2$ is linear with thickness and the SiO$_2$/Si interface thermal resistance is negligible.  Using this calculation, the average thermal conductivity of the SiO$_2$ film in this work is found to be 1.91\,$\pm$\,0.24\,Wm$^{-1}$K$^{-1}$ which is 44\% larger than similar measurements of SiO$_2$ in the literature.\cite{Lee_PRB_1995,Yamane_JAP_2002}  This discrepancy likely arises from a larger metal/SiO$_2$ interface thermal resistance on the 220\,nm SiO$_2$ film caused by the increased surface roughness from the plasma etch.  This effect would increase the measured thermal resistance of the thinner SiO$_2$ film and correspondingly increase the calculated SiO$_2$ thermal conductivity.

In conclusion, we have demonstrated that three-omega thermal-conductivity measurements can be performed with a range of curved heater geometries.  We find any error introduced by wire curvature to be less than one standard deviation of the measurement results when the thermal penetration depth into the substrate is as much as 4.3 times larger than the smallest radii of curvature in the heater.  The Si substrate thermal conductivity measured with curved wire geometries differs from that of straight-wire heater devices by only 0.2\%. The difference in the measured cross-plane thermal resistance between a 285\,nm and 220\,nm SiO$_2$ film varied by as little as 4.3\,m$^2$KGW$^{-1}$.  This result provides needed design criteria for how large a curvature can be included in a heater design when routing heating wires around obstacles on a sample's surface for a three-omega thermal-conductivity measurement.

Measurements are performed with support from US DOE Basic Energy Sciences DE-FG02-03ER46028.  We acknowledge the use of facilities supported by NSF through the UW-Madison MRSEC (DMR-1720415). 

The data that supports the findings of this study are available within the article.

\bibliographystyle{apsrev4-1}
\bibliography{./Bibliography}
\end{document}